\documentclass[fleqn]{revtex4}

\usepackage{amsmath,graphicx}
\usepackage{verbatim,dcolumn}

\begin{document}

\title{Higher-order corrections to spin-spin scalar interactions in HD$^+$ and H$_2^+$}

\author{Jean-Philippe Karr$^{1,2}$, Mohammad Haidar$^1$, Laurent Hilico$^{1,2}$, Zhen-Xiang Zhong$^3$, and Vladimir I. Korobov$^4$}
\affiliation{$^1$Laboratoire Kastler Brossel, Sorbonne Universit\'e, CNRS, ENS-PSL Research University, Coll\`ege de France, 4 place Jussieu, F-75005 Paris, France}
\affiliation{$^2$Universit\'e d'Evry-Val d'Essonne, Universit\'e Paris-Saclay, Boulevard Fran\c cois Mitterrand, F-91000 Evry, France}
\affiliation{$^3$State Key Laboratory of Magnetic Resonance and Atomic and Molecular Physics, Wuhan Institute of Physics and Mathematics, Innovation Academy for Precision Measurement Science and Technology, Chinese Academy of Sciences, Wuhan 430071, China}
\affiliation{$^4$Bogoliubov Laboratory of Theoretical Physics, Joint Institute for Nuclear Research, Dubna 141980, Russia}

\begin{abstract}
The largest hyperfine interaction coefficients in the hydrogen molecular ion HD$^+$, i.e. the electron-proton and electron-deuteron spin-spin scalar interactions, are calculated with estimated uncertainties slightly below 1~ppm. The $(Z\alpha)^2 E_F$ relativistic correction, for which a detailed derivation is presented, QED corrections up to the order $\alpha^3 \ln^2 (\alpha)$ along with an estimate of higher-order terms, and nuclear structure corrections are taken into account. Improved results are also given for the electron-proton interaction coefficient in H$_2^+$, in excellent agreement with RF spectroscopy experiments. In HD$^+$, a 4$\sigma$ difference is found in the hyperfine splitting of the $(v,L)=(0,3) \to (9,3)$ two-photon transition that was recently measured with high precision. The origin of this discrepancy is unknown.
\end{abstract}

\maketitle

\section{Introduction}

The interest of laser spectroscopy of the HD$^+$ molecular ion for the metrology of fundamental constants was pointed out more than forty years ago~\cite{Wing76}. This potential was recently realized in two experiments, in which a one-photon rotational transition~\cite{Alighanbari20} and a two-photon vibrational transition~\cite{Patra20} were measured in the Lamb-Dicke regime, thereby suppressing the first-order Doppler broadening. In these works, a spin-averaged transition frequency was deduced from the measured hyperfine components, with respective accuracies of 13.5 and 2.9 parts per trillion, and compared to theoretical predictions~\cite{Korobov17}, allowing to improve the determination of the proton-electron mass ratio. On the other hand, the experimental data also enables a high-precision investigation of the hyperfine structure of HD$^+$. In the case of the rotational transition~\cite{Alighanbari20}, six hyperfine components were measured with uncertainties of a few tens of~Hz, while in the vibrational transition~\cite{Patra20}, two hyperfine components were measured with uncertainties below 1~kHz.

On the theoretical side, the hyperfine structure of HD$^+$ has been calculated within the Breit-Pauli approximation~\cite{Bakalov06} including the anomalous magnetic moment of the electron, yielding a relative accuracy of order $\alpha^2$. The effective spin Hamiltonian introduced in that work reads
\begin{equation} \label{hdplus-heff}
\begin{array}{@{}l}\displaystyle
H_{\rm eff} =
   E_1 (\mathbf{L}\!\cdot\!\mathbf{s}_e) +
   E_2 (\mathbf{L}\!\cdot\!\mathbf{I}_p) +
   E_3 (\mathbf{L}\!\cdot\!\mathbf{I}_d) +
   E_4 (\mathbf{I}_p\!\cdot\!\mathbf{s}_e) +
   E_5 (\mathbf{I}_d\!\cdot\!\mathbf{s}_e)
\\[1mm]\hspace{11mm}\displaystyle
  + E_6 \Bigl\{
     2 \mathbf{L}^2 (\mathbf{I}_p\!\cdot\!\mathbf{s}_e)
     -3 [(\mathbf{L}\!\cdot\!\mathbf{I}_p) (\mathbf{L}\!\cdot\!\mathbf{s}_e)
         + (\mathbf{L}\!\cdot\!\mathbf{s}_e) (\mathbf{L}\!\cdot\!\mathbf{I}_p)]
  \Bigr\}
\\[1mm]\hspace{11mm}\displaystyle
   + E_7 \Bigl\{
      2 \mathbf{L}^2 (\mathbf{I}_d\!\cdot\!\mathbf{s}_e) -3 [(\mathbf{L}\!\cdot\!\mathbf{I}_d)
      (\mathbf{L}\!\cdot\!\mathbf{s}_e) + (\mathbf{L}\!\cdot\!\mathbf{s}_e) (\mathbf{L}\!\cdot\!\mathbf{I}_d)]
   \Bigr\}
\\[0mm]\hspace{11mm}\displaystyle
  + E_8 \Bigl\{
     2 \mathbf{L}^2 (\mathbf{I}_p\!\cdot\!\mathbf{I}_d) -3 [(\mathbf{L}\!\cdot\!\mathbf{I}_p)
        (\mathbf{L}\!\cdot\!\mathbf{I}_d) + (\mathbf{L}\!\cdot\!\mathbf{I}_p) (\mathbf{L}\!\cdot\!\mathbf{I}_d)]
  \Bigr\}
  + E_9 \Bigl\{
     2 \mathbf{L}^2 \mathbf{I}_d^2
     - \frac{3}{2} (\mathbf{L}\!\cdot\!\mathbf{I}_d) - 3 (\mathbf{L}\!\cdot\!\mathbf{I}_d)^2
  \Bigr\},
\end{array}
\end{equation}
where $\mathbf{s}_e$, $\mathbf{I}_p$, $\mathbf{I}_d$ are the spins of the electron, proton and deuteron, respectively, and $\mathbf{L}$ is the total orbital angular momentum. The largest coefficients are the spin-spin scalar interactions, $E_4 \sim 900$~MHz and $E_5 \sim 140$~MHz in the ground vibrational state, followed by the spin-orbit term $E_1 \sim 30$~MHz and the tensor interaction constants $E_6,E_7$ in the few-MHz range. Other coefficients range from a few kHz to a few tens of kHz. In order to improve the theory further, the first priority is to calculate higher-order corrections to the spin-spin coefficients. This has been done in~\cite{Korobov09,Korobov16} in the case of H$_2^+$; here, we extend this work to the HD$^+$ case~\cite{Alighanbari20,Patra20,Alighanbari18}. In doing so, we make further improvements in the treatment of nuclear structure corrections and higher-order QED corrections, and present extensive numerical results for a range of ro-vibrational states. This will allow a detailed comparison with recent and future experiments when the theoretical precision of next largest coefficients ($E_1,E_6,E_7$) is sufficiently improved. Preliminary results for the $E_1$ coefficient have been obtained in~\cite{Korobov20}.

The paper is organized as follows. In Sec.~\ref{atom-hfs}, we briefly review the theory of the ground-state hyperfine splitting in atomic hydrogen and deuterium. The theory of higher-order corrections to $E_4$ and $E_5$ in HD$^+$ is described in Sec.~\ref{hdplus-theory}. Two contributions that require new calculations are considered separately in the next sections: the relativistic correction of order $(Z\alpha)^2 E_F$, and a vibrational contribution at the next order $\alpha(Z\alpha)^2 E_F$. Finally, numerical results are presented in Sec.\ref{sec-num} and compared with available experimental data.

\section{Ground-state hyperfine structure in the hydrogen and deuterium atoms} \label{atom-hfs}

The ground-state hyperfine splitting of a hydrogenlike atom is given in the nonrelativistic approximation by the so-called Fermi energy~\cite{Bethe77}, which may be written in (SI) frequency units as
\begin{equation}
E_F = \frac{4}{3} \, \mu_0  \frac{\mu_B}{h} \, \left\langle (\mathbf{s}_e\!\cdot\!\boldsymbol{\mu}_M)_{F=I+1/2} - (\mathbf{s}_e\!\cdot\!\boldsymbol{\mu}_M)_{F=I-1/2} \right\rangle \> \langle \delta (\mathbf{r}) \rangle
= \frac{8}{3} Z^3\alpha^2 cR_{\infty} \mu_M \frac{m}{M_p} \frac{2I+1}{2I} \left( 1 + \frac{m}{M} \right)^3. \label{Fermi}
\end{equation}
Here, $\boldsymbol{\mu}_M$ is the nuclear magnetic moment, and $\mu_M$ its value in units of the nuclear Bohr magneton: $\boldsymbol{\mu}_M = \mu_M \mu_N \mathbf{I}/I$ with $\mu_N = |e|\hbar/2M_p$. $\mathbf{I}$ is the nuclear spin, $F$ the total spin quantum number ($\mathbf{F} = \mathbf{I} + \mathbf{s}_e$), and $Z$ the nuclear charge. Finally, $m$, $M_p$ and $M$ are respectively the masses of the electron, proton and nucleus.

QED corrections without recoil terms have been known for some time~\cite{Sapirstein90,Kinoshita96,Eides07,Mondejar10,Mohr16} and may be expressed as
\begin{equation}
\label{hfs-QED}
\begin{array}{@{}l}\displaystyle
\Delta E_{\rm hfs} (QED) = E_F \left\{ a_e + \frac{3}{2} (Z\alpha)^2 + \left( \ln 2 - \frac{5}{2} \right) \alpha (Z \alpha) + \frac{\alpha(Z\alpha)^2}{\pi} \left[ -\frac{8}{3} \ln^2 (Z\alpha) \right. \right.
\\[3mm]\hspace{20mm}\displaystyle
  \left. \left. + \frac{8}{3} \ln (Z\alpha) \left( \ln 4 -\frac{281}{480} \right)  + 16.903\,772\ldots \right] + 0.77099(2) \frac{\alpha^2(Z\alpha)}{\pi} + D^{(4)}(\alpha,Z\alpha) + \ldots \right\} \,,
\end{array}
\end{equation}
where $a_e$ is the electron anomalous magnetic moment. We have kept $Z$ in all expressions in order to identify the origins of different corrections. Corrections at order $\alpha^4 E_F$ have been partially evaluated~\cite{Eides07,Mohr16},
\begin{equation}
\label{hfs-QED-4}
\begin{array}{@{}l}\displaystyle
D^{(4)}(\alpha,Z\alpha) = \frac{17}{8} (Z\alpha)^4 + \alpha(Z\alpha)^3 \left[ \left( -5 \ln 2 + \frac{547}{48} \right) \ln(Z\alpha) -2.102(3) \right]
\\[3mm]\hspace{20mm}\displaystyle
 + \frac{\alpha^2(Z\alpha)^2}{\pi^2} \left[ -\frac{4}{3} \ln^2(Z\alpha) + 1.278 \, 001\ldots\times\ln(Z\alpha) + 10(2.5)\right] - 1.358(1.0) \frac{\alpha^3(Z\alpha)}{\pi^2}.
\end{array}
\end{equation}
Note that the term ($-2.102(3)$) at order $\alpha(Z\alpha)^3$ actually includes corrections of higher order in $Z\alpha$~\cite{Yerokhin08,Yerokhin10,Sunnergren98,Karshenboim00}.

In addition to QED corrections, there are recoil and nuclear structure corrections. In the hydrogen atom case (see~\cite{Sapirstein90,Eides07,Volotka05,Carlson08} for a detailed discussion), these corrections are written as
\begin{equation}
\Delta E_S = \Delta E_Z + \Delta E_R^p + \Delta E_{\rm pol}. \label{struct_H}
\end{equation}
The first and largest term is the Zemach correction~\cite{Zemach56} that reads, including radiative corrections~\cite{Karshenboim97}
\begin{equation}
\Delta E_Z = -2 (Z\alpha) m_r (1 + \delta_Z^{\rm rad}) r_Z E_F \sim -40 \times 10^{-6} \; E_F,
\end{equation}
where $m_r = (m M_p)/(m + M_p)$, $\delta_Z^{\rm rad} = 0.015$, and $r_Z$ is the Zemach radius, a mean radius associated with a convolution of the proton's charge and magnetization distributions,
\begin{equation}
r_Z = \frac{1}{\pi^2} \int \frac{d^3q}{q^4} \left[ 1 - \frac{G_E(-q^2) G_M(-q^2)}{\mu_p} \right].
\end{equation}
$G_E$ and $G_M$ are the proton's electric and magnetic form factors. The second term of Eq.~(\ref{struct_H}) is the recoil correction, where contributions at orders $(Z\alpha)(m/M)E_F$~\cite{Arnowitt53,Newcomb55,Bodwin88,Carlson08}, $(Z\alpha)^2(m/M)E_F$~\cite{Bodwin88} and the $\alpha(Z\alpha)(m/M)E_F$ radiative-recoil correction~\cite{Karshenboim97} add up to around $5.8 \times 10^{-6} \; E_F$~\cite{Faustov02,Carlson08}. Finally, the last term is the proton polarizability correction, evaluations of which yielded the values $1.4(6) \times 10^{-6} \; E_F$~\cite{Faustov02} and $1.88(64) \times 10^{-6} \; E_F$~\cite{Carlson08}.

The deuterium atom case is different, due to the deuteron being a much more weakly bound system than the proton. Nuclear structure corrections are dominated in this case by the deuteron polarizability contribution which amounts to about $240\times10^{-6} \; E_F$~\cite{Friar05}, while the Zemach term contributes at a level of $\sim -100\times10^{-6} \; E_F$~\cite{Friar04}.

In order to get accurate predictions of the HD$^+$ Fermi interaction terms, we will make use of the fact that the total nuclear corrections can be determined phenomenologically with very good accuracy by subtracting the results of the pure QED calculation from the experimental value:
\begin{equation}
\Delta E_{\rm hfs}\mbox{(nucl)} = E_{\rm hfs}\mbox{(exp)} - E_{\rm hfs}\mbox{(QED)}, \label{nucl-corr}
\end{equation}
with $E_{\rm hfs}\mbox{(QED)} = E_F + \Delta E_{\rm hfs}\mbox{(QED)}$. Being due to short-range interactions, the nuclear correction is mainly determined by the squared value of the electronic wavefunction at the nucleus ($\psi(0)^2$) and only very weakly depends on its value at finite distances. This dependence may be neglected without serious loss of accuracy, which allows us to directly plug the nuclear correction, as determined phenomenologically from the experimental atomic hyperfine splitting, in the theory of hydrogen molecular ions, as detailed in the next section.

A summary of QED contributions and the nuclear correction obtained from Eq.~(\ref{nucl-corr}) are shown in Table~\ref{table-h-d-atoms} for both the H and D atoms. For completeness, one should mention additional small corrections not included in $E_{\rm hfs}$(QED), from muonic and hadronic vacuum polarization~\cite{Karshenboim97}, and weak interaction~\cite{Beg74}. $E_{\rm hfs}$(nucl) therefore corresponds to a sum of nuclear corrections and of these contributions.

\begin{table} [h!]
\small
\begin{tabular}{@{\hspace{1mm}}l@{\hspace{1mm}}@{\hspace{1mm}}d@{\hspace{2mm}}@{\hspace{1mm}}d@{\hspace{1mm}}}
\hline\hline
 & \mathrm{H} & \mathrm{D} \\
 \hline
 $E_F$                                          & 1 \, 418 \, 840.093 & 326 \, 967.681 \\
 $a_e \, E_F$                                   &        1 \, 645.361 &        379.169 \\
 $\Delta E_{(Z\alpha)^2}$                       &             113.333 &         26.117 \\
 $\Delta E_{\alpha (Z\alpha)}$                  &            -136.517 &        -31.460 \\
 $\Delta E_{\alpha (Z\alpha)^2 \ln^2(Z\alpha)}$ &             -11.330 &         -2.611 \\
 $\Delta E_{\rm ho}$                            &            1.089(1) &          0.251(1) \\
 \hline
 $E_{\rm hfs}$(QED)                             & 1 \, 420 \, 452.028(1) & 327 \, 339.147(1) \\
 \hline
 $E_{\rm hfs}$(exp)                             & 1 \, 420 \, 405.751 \, 768(1) & 327 \, 384.352 \, 522(2) \\
 \hline
 $E_{\rm hfs}$(nucl)                            &             -46.276 &         45.205 \\
 \hline
 $E_{\rm hfs}$(nucl)$/E_F$ (ppm)                &             -32.616 &         138.256 \\
 \hline\hline
\end{tabular}
\caption{\label{table-h-d-atoms} Contributions (in kHz) to the ground-state hyperfine splitting in the hydrogen and deuterium atoms. The first row is the Fermi energy [Eq.~(\ref{Fermi})], and rows 2-5 are the QED corrections as written in the first line of Eq.~(\ref{hfs-QED}). $\Delta E_{\rm ho}$ corresponds to the terms appearing in the second line of Eq.~(\ref{hfs-QED}); uncertainties take into account the theoretical uncertainties indicated in Eq.~(\ref{hfs-QED-4}), as well as uncertainties of the nuclear magnetic moment values. The experimental values are respectively taken from an adjustment done in~\cite{Karshenboim05} for the hydrogen atom, and from Ref.~\cite{Wineland72} for deuterium.}
\end{table}

\section{Spin-spin scalar interactions in HD$^+$} \label{hdplus-theory}

From here on, we use atomic units. Like in atoms, the electron-proton and electron-deuteron spin-spin scalar interactions are given at the leading orders ($m\alpha^4$ and $m\alpha^5$) by the Fermi term appearing in the Breit-Pauli Hamiltonian, taking into account the anomalous magnetic moment $a_e$ of the electron~\cite{Bakalov06}:
\begin{equation}
H_{ss}^{(0)} = \alpha^2 \frac{8\pi}{3} (1+a_e) \frac{m}{M_p} \left[ \mu_p \delta(\mathbf{r}_p) (\mathbf{s}_e \!\cdot\! \mathbf{I}_p) + \frac{\mu_d}{2} \delta(\mathbf{r}_d) (\mathbf{s}_e \!\cdot\! \mathbf{I}_d) \right].
\end{equation}
The leading contributions to the $E_4$ and $E_5$ hyperfine coefficients (see Eq.~(\ref{hdplus-heff})) are then
\begin{eqnarray}
E_4^{(\rm lo)} &=& (1+a_e) E_4^{(F)} \;,\qquad E_4^{(F)} = \alpha^2 \frac{8\pi}{3} \frac{m}{M_p} \mu_p \langle \delta(\mathbf{r}_p) \rangle, \label{E4lo} \\
E_5^{(\rm lo)} &=& (1+a_e) E_5^{(F)} \;,\qquad E_5^{(F)} = \alpha^2 \frac{4\pi}{3} \frac{m}{M_p} \mu_d \langle \delta(\mathbf{r}_d) \rangle. \label{E5lo}
\end{eqnarray}
This corresponds to the leading-order contribution of Eq.~(\ref{Fermi}), with the first term of Eq.~(\ref{hfs-QED}) included as well, and was the only contribution considered in~\cite{Bakalov06}.

In order to improve the theoretical values of $E_4$ and $E_5$, one should consider higher-order QED corrections and nuclear structure effects, as seen for the atomic case in the previous section. A key point is that the major part of these contributions are state-independent, i.e. contact-type interactions only depending on the value of the squared density of the nonrelativistic wavefunction at the electron-nucleus coalescence point. Such contributions are given by a fixed coefficient taken from the H (respectively D) atom theory regarding corrections to $E_4$ (respectively $E_5$) multiplied by the expectation values of $\delta$-function operators in HD$^+$, which have been already obtained with very high accuracy from variational three-body wavefunctions~\cite{Aznabayev19}. Thus they do not require any new calculations. The most important state-dependent contribution is the relativistic correction of order $(Z\alpha)^2 E_F$ (the second term of Eq.~(\ref{hfs-QED}) in the atomic case, also known as the ``Breit correction''). This term requires an independent calculation, which is presented in the next section. All other terms are included in the form of contact interactions:
\begin{eqnarray}
&&E_{4,5}^{\alpha(Z\alpha)} = \left( \ln 2 - \frac{5}{2} \right) \alpha^2 \; E_{4,5}^{(F)} \label{aZa} \\
&&E_{4,5}^{\alpha (Z\alpha)^2 \ln^2(Z\alpha)} = -\frac{8}{3\pi} \ln^2(\alpha) \, \alpha^3 \; E_{4,5}^{(F)} \label{aZa2} \\
&&E_{4,5}^{\rm (ho)} = 0.767 \times 10^{-6} \; E_{4,5}^{(F)} \label{QED-ho} \\
&&E_{4}^{\rm (nucl)} = -32.616 \times 10^{-6} \; E_{4}^{(F)} \label{nucl-H} \\
&&E_{5}^{\rm (nucl)} = 138.256 \times 10^{-6} \; E_{5}^{(F)} \label{nucl-D}
\end{eqnarray}
The expressions of the first two terms are exact, whereas the next ones are obtained by neglecting the state dependence of the respective contributions. Among the higher-order nonrecoil QED corrections [Eq.~(\ref{QED-ho})], the largest state-dependent term is that of order $\alpha(Z\alpha)^2\ln(Z\alpha)$ (first term in the second line of Eq.~(\ref{hfs-QED})). Among nuclear corrections [Eqs.~(\ref{nucl-H}-\ref{nucl-D})], the only term having a non-negligible state dependence is the recoil correction of order $(Z\alpha)^2(m/M)E_F$~\cite{Bodwin88}, while for other terms the state dependence is much smaller, e.g. they contribute to the specific difference $D_{21} = 8 E_{\rm hfs}(2S) - E_{\rm hfs}(1S)$ at the level of a few~Hz only~\cite{Karshenboim05}. The uncertainty induced by the approximate expressions (\ref{QED-ho}-\ref{nucl-D}) can be estimated as equal to the sum of all state-dependent contributions, leading to a theoretical uncertainty:
\begin{equation}
\Delta E_4 \sim 0.93 \times 10^{-6} \; E_4^{(F)} \;,\qquad \Delta E_5 \sim 0.59 \times 10^{-6} E_5^{(F)}. \label{uncert}
\end{equation}
The difference between the proton and deuteron cases stems from the different magnitude of the $(Z\alpha)^2(m/M)E_F$ recoil correction~\cite{Bodwin88}.

\section{The $(Z\alpha)^2E_F$ relativistic correction} \label{a6rel}

We now derive the ``Breit'' correction to the spin-spin scalar interaction coefficients ($E_4$ and $E_5$) in HD$^+$. In the H$_2^+$ ion, the corresponding contribution was calculated in~\cite{Korobov09,Korobov16}. The derivation presented here for HD$^+$ is similar in spirit, but differs in the details due the existence of two separate interaction constants instead of a global interaction between the electron spin and the total nuclear spin.

We use the adiabatic approximation, and calculate the correction to the bound electron. The nonrelativistic electronic Hamiltonian is
\begin{equation}\label{HDplus_nonrel}
H_0 = \frac{p^2}{2m}+V,\qquad V=-\frac{Z_1}{r_1}-\frac{Z_2}{r_2},
\end{equation}
where $\mathbf{p}$ is the electron's impulse operator, $Z_1,Z_2$ the nuclear charges and $r_1,r_2$ the distances between the electron and the nuclei. In what follows we will assume that $Z_1\!=\!Z_2\!=\!Z$. We respectively denote by $E_0$ and $\psi_0$ the nonrelativistic energy and wavefunction (in our numerical calculations, we will consider only the ground $1s\sigma$ electronic state).

%and $\boldsymbol{\mu}_p\!=\!Z(\mu_p/m_p)\mathbf{I}_p$ and $\boldsymbol{\mu}_d\!=\!Z(\mu_d/2m_p)\mathbf{I}_d$, where $\mu_p$ and $\mu_d$ are magnetic moments in nuclear magnetons and there values are:
%$\mu_p=2.792847356(23)\mu_N$ and $\mu_d=0.8574382308(72)\mu_N$.

The first step consists in deriving the effective potentials of order $m\alpha^6(m/M)$ by the NRQED approach. This was done in~\cite{Korobov09}, and these potentials were later rederived in~\cite{Korobov20} along with other spin-dependent interactions at order $m\alpha^6$. Here, we use the notations of Ref.~\cite{Korobov09}. The potentials contributing to the spin-spin scalar interactions are the following (Eq.~(42) of~\cite{Korobov09}, see also Eq.~(33) of~\cite{Korobov20}):
\begin{equation}\label{eq:potentials468}
\begin{array}{@{}l}\displaystyle
\mathcal{V}_4 =
   -\alpha^4\frac{1}{4m^3}
      \left\{
         p^2,
         \left[
            \frac{8\pi}{3}\mathbf{s}_e\!\cdot\!\boldsymbol{\mu}_i
                                                \delta(\mathbf{r}_i)
            -\frac{r_i^2\mathbf{s}_e\!\cdot\!\boldsymbol{\mu}_i
                 \!-\!3(\mathbf{s}_e\!\cdot\!\mathbf{r}_i)
                       (\boldsymbol{\mu}_i\!\cdot\!\mathbf{r}_i)}{r_i^5}
         \right]
      \right\},
\\[4mm]\displaystyle
\mathcal{V}_6=
    \alpha^4\frac{Z}{6m^2}
    \left[
    \frac{2(\mathbf{r}_1\!\cdot\!\mathbf{r}_2)(\mathbf{s}_e\!\cdot\!\boldsymbol{\mu}_I)}
                                                         {r_1^3r_2^3}
    +\frac{(\mathbf{r}_1\!\cdot\!\mathbf{r}_2)(\mathbf{s}_e\!\cdot\!\boldsymbol{\mu}_I)
          \!-\!3(\mathbf{r}_1\!\cdot\!\mathbf{s}_e)(\mathbf{r}_2\!\cdot\!\boldsymbol{\mu}_2)
          \!-\!3(\mathbf{r}_2\!\cdot\!\mathbf{s}_e)(\mathbf{r}_1\!\cdot\!\boldsymbol{\mu}_1)}
                                                       {r_1^3r_2^3}
    \right],
\\[4mm]\displaystyle
\mathcal{V}_8 = \alpha^4\frac{Z}{6m^2}\;
    \left[
       \frac{2(\mathbf{s}_e\!\cdot\!\boldsymbol{\mu}_i)}{r_i^4}
       +\frac{r_i^2(\mathbf{s}_e\!\cdot\!\boldsymbol{\mu}_i)
             \!-\!3(\mathbf{r}_i\!\cdot\!\mathbf{s}_e)(\mathbf{r}_i\!\cdot\!\boldsymbol{\mu}_i)}
                                                             {r_i^6}
    \right].
\end{array}
\end{equation}
Here, $\{X,Y\} = XY + YX$. $\boldsymbol{\mu}_1,\boldsymbol{\mu}_2$ are the nuclear magnetic moments ($\boldsymbol{\mu}_1 \equiv \boldsymbol{\mu}_p$, $\boldsymbol{\mu}_2 \equiv \boldsymbol{\mu}_d$), and $\boldsymbol{\mu}_I=\boldsymbol{\mu}_1+\boldsymbol{\mu}_2$. In each line, the first and second term contribute to scalar and tensor interactions, respectively. Keeping only the terms contributing to scalar interactions, we arrive at the total effective Hamiltonian:
\begin{equation}
\begin{array}{@{}r@{\;}l}
H_s^{(6)} & \displaystyle
 = \alpha^4
   \left[
      -\frac{1}{6m^3}\Bigl\{p^2,4\pi\delta(\mathbf{r}_1)\Bigr\}
      +\frac{Z}{3m^2}
      \left(
         \frac{1}{r_1^4}
         +\frac{\mathbf{r}_1\!\cdot\!\mathbf{r}_2}{r_1^3r_2^3}
         \right)
   \right]\mathbf{s}_e\!\cdot\!\boldsymbol{\mu}_1
\\[4mm] & \displaystyle\hspace{20mm}
   +\alpha^4\left[
      -\frac{1}{6m^3}\Bigl\{p^2,4\pi\delta(\mathbf{r}_2)\Bigr\}
      +\frac{Z}{3m^2}
      \left(
         \frac{1}{r_2^4}
         +\frac{\mathbf{r}_1\!\cdot\!\mathbf{r}_2}{r_1^3r_2^3}
      \right)
   \right]\mathbf{s}_e\!\cdot\!\boldsymbol{\mu}_2 \>.
\end{array}
\end{equation}
From now on, we focus on the $\mathbf{s}_e\!\cdot\!\boldsymbol{\mu}_1$ interaction term, calculations for the other term being identical. It may be rewritten as
\begin{equation} \label{Heff6}
H_{s1}^{(6)}
 = \frac{\alpha^4}{3Zm^2}
   \left[
      -\frac{1}{2m}\left\{p^2,\rho_1\right\}
      + \boldsymbol{\mathcal{E}}_1\!\cdot\!\boldsymbol{\mathcal{E}}
   \right]\mathbf{s}_e\!\cdot\!\boldsymbol{\mu}_1,
\end{equation}
with the definitions: $V_i = -Z/r_i$, $4\pi \rho_i = \Delta V_i$, $\boldsymbol{\mathcal{E}}_i = - \boldsymbol{\nabla}V_i$ ($i = 1,2$), and $\rho = \rho_1 + \rho_2$,  $\boldsymbol{\mathcal{E}} = \boldsymbol{\mathcal{E}}_1 + \boldsymbol{\mathcal{E}}_2$.

According to the nonrelativistic perturbation theory, the total energy correction of order $m\alpha^6(m/M)$ to the $\mathbf{s}_e\!\cdot\!\boldsymbol{\mu}_1$ interaction is given by
\begin{equation} \label{DeltaE6-total}
\Delta E^{(6)}_{s1} = \left\langle H_{s1}^{(6)} \right\rangle +    2\alpha^4  \left\langle
      H_B^{(2)} Q(E_0-H_0)^{-1}Q H_{ss1}^{(1)}
   \right\rangle.
\end{equation}
Here, we have omitted a second-order perturbation term induced by the electronic spin-orbit and nuclear spin-orbit interactions (respectively denoted by $H_{so}$ and $H_{so_N}$ in Eq.~(28) of Ref.~\cite{Korobov20}), which is negligibly small for $\sigma$ electronic states. The brackets denote an expectation value over the nonrelativistic electronic wavefunction $\psi_0$, and $Q$ is a projection operator on a subspace orthogonal to $\psi_0$. $H_{ss1}^{(1)}$ is the leading-order Fermi interaction, and $H_B^{(2)}$ is the spin-independent part of the Breit-Pauli Hamiltonian accounting for leading-order relativistic corrections to the bound electron:
\begin{equation}
\begin{array}{@{}l}\displaystyle
H_B^{(2)}=-\frac{p^4}{8m^3}
          +\frac{Z}{8m^2}\,
             4\pi\Bigl(\delta(\mathbf{r}_1)+\delta(\mathbf{r}_2)\Bigr),
\\[3.5mm]\displaystyle
H_{ss1}^{(1)} = \frac{2}{3m} H_{B1}^{(1)} \, \mathbf{s}_e \!\cdot\! \boldsymbol{\mu}_1,
\end{array}
\end{equation}
here $H_{B1}^{(1)}=4\pi Z\delta(\mathbf{r}_1)$. Both terms in Eq.~(\ref{DeltaE6-total}) are divergent, but their sum is finite. They need to be transformed in order to separate and cancel divergent terms, as was done in~\cite{Korobov09} for H$_2^+$.

The first-order term can be transformed using the relationship $p^2\Psi_0\!=\!2m(E_0\!-\!V)\Psi_0$, commutation relations, and integration by parts (see Appendix~\ref{div-op} for details). One gets, using Eq.~(\ref{c4}):
\begin{equation}
\begin{array}{@{}l}
\displaystyle
\left\langle H_{s1}^{(6)} \right\rangle
 = \frac{\alpha^4}{3Zm^2}
   \Bigl[
      -\left\langle
         \boldsymbol{\mathcal{E}}_1 \!\cdot\! \boldsymbol{\mathcal{E}}
      \right\rangle
      +4m\left\langle V_1V^2 \right\rangle
      -4mE_0\left\langle V_1V \right\rangle
      +2\left\langle \mathbf{p}V_1V\mathbf{p} \right\rangle
\\[2.5mm]\hspace{50mm}
      +4\pi\left\langle V_2\rho_1\!-\!V_1\rho_2 \right\rangle
      -8\pi E_0\left\langle \rho_1 \right\rangle
   \Bigr]\left\langle \mathbf{s}_e\!\cdot\!\boldsymbol{\mu}_1 \right\rangle.
\end{array}
\end{equation}

Let us now consider the second-order term. We introduce the first-order perturbation wavefunction $\Psi_{B1}^{(1)}$, solution of the equation
\begin{equation}\label{HDplus_WFdelta}
(E_0-H_0) \Psi_{B1}^{(1)} = Q H_{B1}^{(1)} \Psi_0.
\end{equation}
This wavefunction behaves like $1/r_1$ in the limit $r_1 \to 0$. The $1/r_1$ singularity can be separated by setting
\[
\Psi_{B1}^{(1)} = -\frac{2Zm}{r_1}\Psi_0
               +\tilde{\Psi}_{B1}^{(1)}
             = U_1\Psi_0+\tilde{\Psi}_{B1}^{(1)},
\qquad
U_1 = 2mV_1,
\]
where $\tilde{\Psi}_{B1}^{(1)}$ is a less singular function, behaving like $\ln{r_1}$ at $r_1 \to 0$, and is a solution of the equation
\[
(E_0-H_0)\tilde{\Psi}_{B1}^{(1)} =
   \left(H'^{(1)}_{B1}
   -\left\langle H'^{(1)}_{B1} \right\rangle\right)\Psi_0 \, ,
\qquad
H'^{(1)}_{B1} =-(E_0-H_0)U_1-U_1(E_0-H_0)+H_{B1}^{(1)}
\]
Similarly, for ${H_{B\!}}^{(2)}$, we introduce the first-order wavefunction $\Psi_{B}^{(2)}$:
\begin{equation}\label{Hplus_WFbreit}
(E_0-H_0)\Psi_B^{(2)} = Q H_B^{(2)} \Psi_0,
\end{equation}
and separate its $1/r_1$ and $1/r_2$ singularities:
\[
\Psi_B^{(2)} = \frac{Z}{4m}
                  \left[-\frac{1}{r_1}-\frac{1}{r_2}\right]\Psi_0(r)
                               +\tilde{\Psi}_B^{(2)}
             = U_2\Psi_0+\tilde{\Psi}_B^{(2)},
\qquad
U_2 = -\frac{1}{4m}V.
\]
Here $\tilde{\Psi}_B^{(2)}$ is a solution of the equation
\[
(E_0-H_0)\tilde{\Psi}_B^{(2)} =
   \left({H'^{(2)}_B}
   -\left\langle{H'^{(2)}_B}\right\rangle\right)\Psi_0 \, ,
\qquad
H'^{(2)}_B = -(E_0-H_0)U_2-U_2(E_0-H_0)+H_B^{(2)}
\]
Then, the divergent part of the second-order term can be separated as follows:
\[
\begin{array}{@{}r@{\;}l}
\Delta E_{A1} & = \displaystyle
   \alpha^4\frac{4}{3m}
   \left\langle
      \Psi_0\Big|H_B^{(2)}Q(E_0-H_0)^{-1}QH_{B1}^{(1)}\Big|\Psi_0
   \right\rangle
   \left\langle \mathbf{s}_e \!\cdot\! \boldsymbol{\mu}_1 \right\rangle
\\[3mm] & \displaystyle
 = \alpha^4\frac{4}{3m}
\Bigl(
   \left\langle
      \Psi_0\Big|\left(H_B^{(2)}
         -\left\langle H_B^{(2)}\right\rangle\right)U_1\Big|\Psi_0
   \right\rangle+
   \left\langle
      \Psi_0\Big|U_2\left(H'^{(1)}_{B1}
      -\left\langle H'^{(1)}_{B1}\right\rangle\right)\Big|\Psi_0
   \right\rangle
\\[3mm] & \displaystyle \hspace{45mm}
   +\left\langle
      \Psi_0\Big|H'^{(2)}_{B}Q(E_0-H_0)^{-1}QH'^{(1)}_{B1}\Big|\Psi_0
   \right\rangle
\Bigr) \left\langle \mathbf{s}_e \!\cdot\! \boldsymbol{\mu}_1 \right\rangle
\end{array}
\]
In this expression, the last term is finite. The divergences are located in the first two terms, which can be written as the expectation value of an effective Hamiltonian:
\begin{equation} \label{2nd-order-div}
\begin{array}{@{}r@{\;}l}
\displaystyle
H'^{(6)}_{s1} & \displaystyle
 = \alpha^4\frac{2}{3m}
   \biggl\{
      \left( H_B^{(2)}U_1+U_1H_B^{(2)}\right)
      +\left( H_{B1}^{(1)}U_2+U_2H_{B1}^{(1)}\right)
      -2\left\langle H_B^{(2)}\right\rangle U_1
      -2\left\langle H_{B1}^{(1)}\right\rangle U_2
\\[2mm] & \hspace{60mm}
      -U_1(E_0-H_0)U_2-U_2(E_0-H_0)U_1
   \biggr\}
   \left( \mathbf{s}_e \!\cdot\! \boldsymbol{\mu}_1 \right)
\\[2mm] & \displaystyle
 = \alpha^4\frac{2}{3Zm}
   \biggl[
      -\frac{p^4V_1\!+\!V_1p^4}{4m^2}
      +\frac{4\pi Z\Bigl[\rho_2 V_1\!-\!\rho_1 V_2\Bigr]}{2m}
      -\frac{(V_1p^2V\!+\!Vp^2V_1)}{4m}
\\[2mm] & \displaystyle\hspace{30mm}
      -V_1V^2+E_0V_1V
      -4m\left\langle H_B^{(2)}\right\rangle V_1
      +\frac{\bigl\langle H_{B1}^{(1)}\bigr\rangle V}{2m}
   \biggr]
   \left( \mathbf{s}_e \!\cdot\! \boldsymbol{\mu}_1 \right)
\end{array}
\end{equation}
Using Eqs.~(\ref{B4}) and~(\ref{c3}) (see Appendix~\ref{div-op}), the expectation value of $H'^{(6)}_{s1}$ can be transformed to:
\begin{equation}
\begin{array}{@{}r@{\;}l}
\displaystyle
\left\langle H'^{(6)}_{s1} \right\rangle & \displaystyle
 = \alpha^4\frac{1}{3Zm^2}
   \biggl[
      \left\langle
         \boldsymbol{\mathcal{E}}_1 \!\cdot\! \boldsymbol{\mathcal{E}}
      \right\rangle
      -4m\left\langle V_1V^2 \right\rangle
      +8mE_0\left\langle V_1V \right\rangle
      -4\pi\left\langle V_2\rho_1\!-\!V_1\rho_2 \right\rangle
\\[1mm] & \hspace{25mm} \displaystyle
      -4mE_0^2\left\langle V_1 \right\rangle
      -8m^2\left\langle H_B^{(2)}\right\rangle \langle V_1 \rangle
      +\bigl\langle H_{B1}^{(1)}\bigr\rangle \langle V \rangle
   \biggr]
   \left\langle \mathbf{s}_e \!\cdot\! \boldsymbol{\mu}_1 \right\rangle
\end{array}
\end{equation}
Finally, the total correction of order $m\alpha^6(m/M)$ to the $E_4$ coefficient is given by
\begin{equation}\label{Za2-tot}
\Delta E^{(6)}_{4} = \Delta E'^{(6)}_{A1} + \Delta E'^{(6)}_{B1},
\end{equation}
\vspace*{-3mm}
\addtocounter{equation}{-1}
\begin{subequations}
\begin{equation}\label{Za2-1}
\Delta E'^{(6)}_{A1} =
   \alpha^4\frac{4}{3} \frac{m}{M_p} \mu_p \,
   \left\langle
      \Psi_0\Big| H'^{(2)}_B Q(E_0-H_0)^{-1}Q H'^{(1)}_{B1} \Big|\Psi_0
   \right\rangle,
\end{equation}
\vspace*{-4mm}
\begin{equation}\label{Za2-2}
\begin{array}{@{}l}\displaystyle
\Delta E'^{(6)}_{B1} =
   \alpha^4\frac{2}{3Z} \frac{m}{M_p} \mu_p \,
   \biggl[
      \left\langle \mathbf{p}V_1V\mathbf{p} \right\rangle
      +2mE_0\left\langle V_1V \right\rangle
      -2mE_0^2\left\langle V_1 \right\rangle
\\[2.5mm]\displaystyle\hspace{40mm}
      -4\pi\>E_0 \left\langle \rho_1 \right\rangle
      -4m^2\left\langle H_B^{(2)}\right\rangle \langle V_1 \rangle
      +\frac{1}{2}\bigl\langle H_{B1}^{(1)}\bigr\rangle \langle V \rangle
   \biggr].
\end{array}
\end{equation}
\end{subequations}
Both the second-order term and the first-order term, in which the divergent terms proportional to $\left\langle \boldsymbol{\mathcal{E}}_1 \!\cdot\! \boldsymbol{\mathcal{E}} \right\rangle$ and to $\left\langle V_1V^2 \right\rangle$ have been cancelled out in the sum $\left\langle H^{(6)}_{s1} \right\rangle + \left\langle H'^{(6)}_{s1} \right\rangle$, are now finite. Expressions for the $E_5$ coefficient are identical, excepted that the prefactor $(m/M_p)\mu_p$ is replaced by $(m/(2M_p)) \mu_d$.

Comparing our final result~(\ref{Za2-1})-(\ref{Za2-2}) with the expression obtained in H$_2^+$, Eqs.~(49)-(50) of Ref.~\cite{Korobov09}, it is easily seen that they are equivalent under the assumption that the electronic wavefunction is symmetric with respect to the exchange of nuclei, which is the case in the standard adiabatic approximation that we will use here~\cite{Wolniewicz80}. Indeed, under this assumption the following equalities hold:
\[
\left\langle \mathbf{p}V_1V\mathbf{p} \right\rangle = \frac{1}{2} \left\langle \mathbf{p} V^2 \mathbf{p} \right\rangle, \qquad
\left\langle V_1V \right\rangle = \frac{1}{2} \left\langle V^2 \right\rangle,
\qquad
\left\langle V_1 \right\rangle = \frac{1}{2} \left\langle V \right\rangle,
\qquad
\left\langle \rho_1 \right\rangle = \left\langle \rho_2 \right\rangle.
\]
In the adiabatic framework, corrections to ro-vibrational energy levels are obtained by averaging the correction curve $\Delta E^{(6)}_{4,5}(R)$ over the adiabatic vibrational wave functions $\chi_{v,L}(R)$. The second-order perturbation term induced by the leading-order Fermi interaction and the spin-independent Breit-Pauli Hamiltonian requires specific attention. The correction written in the second term of Eq.~(\ref{DeltaE6-total}) accounts for the perturbation of the electronic part of the wave function only, and one should also take into account the perturbation of the vibrational wavefunction caused by the shift of the potential energy curve ~\cite{Korobov16}. The total correction is
\begin{equation}\label{E6-tot}
\Delta E_4^{(Z\alpha)^2} (v,L) = \Delta E_4^{(Z\alpha)^2({\rm el})} (v,L) + \Delta E_4^{(Z\alpha)^2({\rm vb})} (v,L),
\end{equation}
\vspace*{-3mm}
\addtocounter{equation}{-1}
\begin{subequations}
\begin{equation}\label{E6-el}
\Delta E_4^{(Z\alpha)^2({\rm el})} (v,L) = \langle \chi_{v,L} | \Delta E^{(6)}_{4}(R) | \chi_{v,L} \rangle,
\end{equation}
\vspace*{-4mm}
\begin{equation}\label{E6-vb}
\Delta E_4^{(Z\alpha)^2({\rm vb})} (v,L) = 2 \alpha^4 \langle \chi_{v,L} | E_B^{(2)}(R) Q' (E_0 - H_{\rm vb})^{-1} Q' E_{ss1}(R) | \chi_{v,L} \rangle.
\end{equation}
\end{subequations}
In the last line, $E_{ss1}(R) = \langle H_{ss1} \rangle$, $E_B^{(2)}(R) = \langle H_{B}^{(2)} \rangle$, $Q'$ is a projection operator onto a subspace orthogonal to $\chi_{v,L}$, and $H_{\rm vb}$ is the nuclear radial Hamiltonian~\cite{Wolniewicz80}
\[
H_{\rm vb} = -\frac{\Delta_{\mathbf{R}}} {2 \mu} + U(R) + \frac{L(L+1)}{2 \mu R^2},
\qquad
U(R) = E_0(R) + \frac{Z^2}{R} - \frac{\langle \Delta_{\mathbf{r}} \rangle}{8 \mu} - \frac{\langle \Delta_{\mathbf{R}} \rangle}{2 \mu},
\]
where $\mu = m_p m_d/(m_p + m_d)$, and $\mathbf{r}$ denotes the electronic coordinates.

\section{Vibrational correction of order $\alpha(Z\alpha)^2E_F$ }

There is one last contribution that should be included in our theory. As illustrated in the preceding section, in a molecular system, second-order correction terms consist of an ``electronic'' contribution (the second term of Eq.~(\ref{DeltaE6-total})) and a ``vibrational'' one (Eq.~(\ref{E6-vb})). Our estimate of higher-order corrections in Eq.~(\ref{QED-ho}) only includes the electronic part, so that the vibrational part must be included separately. One such contribution is significant at the level of the theoretical uncertainties~(\ref{uncert}), namely the $\alpha(Z\alpha)^2E_F$-order term induced by the leading-order Fermi interaction and leading-order radiative corrections:
\begin{equation}
\Delta E_4^{\alpha(Z\alpha)^2({\rm vb})} (v,L) = 2 \alpha^5 \langle \chi_{v,L} | E_{\rm rad}(R) Q' (E_0 - H_{\rm vb})^{-1} Q' E_{ss1}(R) | \chi_{v,L} \rangle, \label{E7-vb}
\end{equation}
with
\begin{equation}
E_{\rm rad}(R) = \frac{4}{3} \left[ \ln \frac{1}{\alpha^2} - \beta(R) + \frac{5}{6} - \frac{1}{5} \right] \, Z \langle \delta(\mathbf{r}_1) + \delta(\mathbf{r}_2) \rangle.
\end{equation}
$\beta(R)$ is the nonrelativistic Bethe logarithm for the bound electron. Its values as a function of $R$ can be found in the Supplemental Material of~\cite{Korobov13}.

Before discussing numerical results, it is worth summarizing the improvements brought in our present treatment with respect to that presented in Refs.~\cite{Korobov09,Korobov16} in H$_2^+$, and used in HD$^+$ in Refs.~\cite{Alighanbari20,Patra20,Alighanbari18}:
\begin{itemize}
{\item Nuclear corrections are determined phenomenologically from the difference between the experimental ground-state hyperfine splitting in the H and D atoms and the QED prediction [Eq.~(\ref{nucl-corr})]. Note that the treatment of Ref.~\cite{Korobov09}, where nuclear correction terms were calculated individually, is actually similar in spirit since a value of the proton's Zemach radius deduced from the experimental H-atom ground-state hyperfine splitting was used in that work~\cite{Volotka05}. The more self-consistent approach used here only leads to small differences in $E_4$ and $E_5$ ($\sim 100-200$~Hz).}
{\item More importantly, we take into account an estimate of higher-order nonrecoil QED corrections [Eq.~(\ref{QED-ho})], which includes as well the ``vibrational'' contribution of order $\alpha(Z\alpha)^2E_F$ [Eq.~(\ref{E7-vb})]. In addition, the sum of state-dependent corrections gives an estimate of the theoretical uncertainty.}
\end{itemize}

\section{Numerical results and discussion} \label{sec-num}

We report here the results of calculations of the $E_4$ and $E_5$ hyperfine coefficients in HD$^+$ and of the $b_F$ spin-spin coefficient in H$_2^+$~\cite{Korobov09,Korobov16}, for a range of ro-vibrational states. For the leading-order contribution [Eqs.~(\ref{E4lo})-(\ref{E5lo})], as well as QED and nuclear corrections [Eqs.~(\ref{aZa})-(\ref{nucl-D})], expectation values of $\delta$-function operators are taken from Ref.~\cite{Aznabayev19}. The potential curve corresponding to the $(Z\alpha)^2 E_F$ relativistic correction, $E_{s1}^{(6)}(R)$ [Eqs.~(\ref{Za2-tot})-(\ref{Za2-2})] has been shown to be identical to that obtained in the H$_2^+$ case~\cite{Korobov09} in the adiabatic approximation used here. To calculate corrections to rovibrational levels [Eqs.~(\ref{E6-tot})-(\ref{E6-vb}) and~(\ref{E7-vb})], adiabatic vibrational wavefunctions are obtained by solving numerically the radial Schr\"odinger equation for the nuclear motion.

In H$_2^+$, the results presented here represent a slight improvement with respect to those of Ref.~\cite{Korobov16}. As explained above, the main improvement is that (estimated) higher-order QED corrections are taken into account through the term $b_F^{\rm (ho)} = 0.767\!\times\!10^{-6} \; b_F^{(F)}$ (where $b_F^{(F)}$ is the Fermi value of $b_F$), along with the vibrational contribution [Eq.~(\ref{E7-vb})]. As shown in Table~\ref{table-bF}, the excellent agreement with experiments reported in~\cite{Korobov16} is not significantly altered by the inclusion of higher-order QED effects. More extensive results for the range $(L=1,3, 0 \leq v \leq 10)$ are reported in Table~\ref{table-bF-complete}.

\begin{table} [h!]
\begin{tabular}{@{\hspace{2mm}}c@{\hspace{2mm}}@{\hspace{2mm}}c@{\hspace{2mm}}@{\hspace{2mm}}c@{\hspace{2mm}}@{\hspace{2mm}}c@{\hspace{2mm}}}
\hline \hline
$v$ & ~\cite{Korobov16} & This work & Experiment~\cite{Jefferts69} \\
\hline
4   &  836.7294  &  836.7287(8)  &  836.7292(8)  \\
5   &  819.2272  &  819.2267(8)  &  819.2273(8)  \\
6   &  803.1750  &  803.1745(7)  &  803.1751(8)  \\
7   &  788.5079  &  788.5075(7)  &  788.5079(8)  \\
8   &  775.1714  &  775.1712(7)  &  775.1720(8)  \\
\hline \hline
\end{tabular}
\caption{Theoretical and experimental values (in MHz) of the spin-spin scalar interaction coefficient $b_F$ for a few rovibrational states of $H_2^+$. The rotational quantum number is $L=1$. \label{table-bF}}
\end{table}

\begin{table} [h!]
\small
\begin{tabular}{@{\hspace{2mm}}c@{\hspace{2mm}}@{\hspace{2mm}}c@{\hspace{2mm}}@{\hspace{2mm}}c@{\hspace{2mm}}}
\hline \hline
$v$  & $b_F(L=1,v)$ & $b_F(L=3,v)$ \\
\hline
 0  & 922.9301(9)  & 917.5297(9)  \\
 1  & 898.7493(8)  & 893.6950(8)  \\
 2  & 876.3961(8)  & 871.6699(8)  \\
 3  & 855.7560(8)  & 851.3422(8)  \\
 4  & 836.7287(8)  & 832.6136(8)  \\
 5  & 819.2267(8)  & 815.3988(8)  \\
 6  & 803.1745(7)  & 799.6241(7)  \\
 7  & 788.5075(7)  & 785.2269(7)  \\
 8  & 775.1712(7)  & 772.1546(7)  \\
 9  & 763.1211(7)  & 760.3644(7)  \\
10  & 752.3219(7)  & 749.8233(7)  \\
\hline \hline
\end{tabular}
\caption{Values (in MHz) of the spin-spin scalar interaction coefficient $b_F$ for rovibrational states $(L,v)$ of H$_2^+$. \label{table-bF-complete}}
\end{table}

\begin{table} [h!]
\small
\begin{tabular}{@{\hspace{1mm}}c@{\hspace{3mm}}c@{\hspace{2mm}}c@{\hspace{4mm}}c@{\hspace{4mm}}c@{\hspace{4mm}}c
                @{\hspace{4mm}}c@{\hspace{4mm}}c@{\hspace{3mm}}c@{\hspace{3mm}}c@{\hspace{4mm}}c@{\hspace{4mm}}c
                @{\hspace{1mm}}}
\hline \hline
  & $L$  &  $v$  & lo &  $(Z\alpha)^2$  & $\alpha(Z\alpha)$ &  $\alpha(Z\alpha)^2$  &  ho   &   $\alpha(Z\alpha)^2({\rm vb})$   &   nucl.  &  This work  &    \\
\hline
      &  0  &  0  & 925.4559  & 0.0669  & $-$0.0889  & $-$0.0074  & 0.0007  & $-$0.0029  & $-$0.0301  & 925.3942(9) & 925.39588~\cite{Alighanbari20} \\
      &  1  &  0  & 924.6295  & 0.0669  & $-$0.0889  & $-$0.0074  & 0.0007  & $-$0.0029  & $-$0.0301  & 924.5677(9) & 924.56943~\cite{Alighanbari20} \\
$E_4$ &  3  &  0  & 920.5415  & 0.0665  & $-$0.0885  & $-$0.0073  & 0.0007  & $-$0.0029  & $-$0.0300  & 920.4800(9) & 920.48165~\cite{Patra20}\\
      &  3  &  9  & 775.7556  & 0.0572  & $-$0.0746  & $-$0.0062  & 0.0006  & $-$0.0012  & $-$0.0253  & 775.7061(7) & 775.70633~\cite{Patra20}\\
      &  1  &  6  & 816.7692  & 0.0597  & $-$0.0785  & $-$0.0065  & 0.0006  & $-$0.0018  & $-$0.0266  & 816.7161(8) & \\
\hline
      &  0  &  0  & 142.27278 & 0.01027 & $-$0.01367 & $-$0.00113 & 0.00011 & $-$0.00045 & 0.01965  & 142.28756(8) & 142.28781~\cite{Alighanbari20} \\
      &  1  &  0  & 142.14591 & 0.01026 & $-$0.01366 & $-$0.00113 & 0.00011 & $-$0.00045 & 0.01963  & 142.16067(8) & 142.16092~\cite{Alighanbari20} \\
$E_5$ &  3  &  0  & 141.51840 & 0.01020 & $-$0.01360 & $-$0.00113 & 0.00011 & $-$0.00044 & 0.01954  & 141.53307(8) & 141.53332~\cite{Patra20}\\
      &  3  &  9  & 119.41918 & 0.00879 & $-$0.01148 & $-$0.00095 & 0.00009 & $-$0.00019 & 0.01649  & 119.43193(7) & 119.43196~\cite{Patra20}\\
      &  1  &  6  & 125.64226 & 0.00916 & $-$0.01207 & $-$0.00100 & 0.00010 & $-$0.00027 & 0.01735  & 125.65551(7) & \\
\hline \hline
\end{tabular}
\caption{Contributions to $E_4$ and $E_5$ (in MHz) for a few rovibrational states $(L,v)$ of HD$^+$ (columns 3-9). Our final theoretical values are given in column 10. The theoretical values given in~\cite{Alighanbari20,Patra20} are shown in the last column for comparison.\label{table-E4-E5-exp}}
\end{table}

In HD$^+$, all contributions to $E_4$ and $E_5$ are shown in detail in Table~\ref{table-E4-E5-exp} for a few rovibrational states probed in high-precision experiments~\cite{Alighanbari20,Patra20,Zhong15}, while complete results for a range of states $(0 \leq L \leq 4,0 \leq v \leq 10)$ are given in Table~\ref{table-E4-E5}. Inspection of Table~\ref{table-E4-E5-exp} reveals that our values of $E_4$ ($E_5$) are smaller than those given in~\cite{Alighanbari20,Patra20} by about 1.7~kHz (0.25~kHz) for $v=0$ states, while differences are much smaller for $v=9$ ($\sim 0.2$~kHz for $E_4$ and $0.03$~kHz for $E_5$), due to the smaller value of the $\alpha(Z\alpha)^2 E_F$ vibrational correction. This shifts the measured hyperfine components of the $L=0 \to 1$ rotational transition~\cite{Alighanbari20} by only a few tens of~Hz, which does not significantly change the level of agreement between theory and experiment. This is due to a strong cancellation effect, as transition frequencies essentially depend on the differences $E_4 (v\!=\!0,L\!=\!1)\!-\!E_4(v\!=\!0,L\!=\!0)$ and $E_5 (v\!=\!0,L\!=\!1)\!-\!E_5(v\!=\!0,L\!=\!0)$. The rotational transition is thus not a stringent test for the theory of spin-spin scalar interactions, and is much more sensitive to the spin-orbit and tensor coefficients ($E_1,E_6,E_7$) in the $L=1$ state. A more detailed analysis requires calculation of higher-order corrections to these coefficients, which is currently in progress~\cite{Korobov20}.

In the $(L,v)=(3,0) \to (3,9)$ two-photon transition~\cite{Patra20}, the additional contributions included in the present work decrease the theoretical value of the separation between the two measured hyperfine components by about 1.5~kHz, down to
\begin{equation}
f_{\rm hfs,theo} = 178.2462(18)~\mbox{MHz},
\end{equation}
to be compared with the experimental value
\begin{equation}
f_{\rm hfs,exp} = 178.2544(9)~\mbox{MHz}.
\end{equation}
The values of the other hyperfine coefficients can be found in the Supplementary Materials of Ref.~\cite{Patra20}, excepted for $E_1$ where we used the value of Ref.~\cite{Korobov20}. To estimate the theoretical uncertainty, we have assumed an uncertainty of 400~Hz for the $E_1$ coefficient~\cite{Korobov20}, and a relative uncertainty of 10$^{-4}$ for the other (smaller) coefficients. The difference between theory and experiment amounts to 8.2~kHz or 4.1 combined standard deviations, whereas in H$_2^+$, excellent agreement, within the 0.8~kHz experimental error bar, is obtained with the same theoretical ingredients.

The origin of this discrepancy is unknown. It is unlikely that it is due to an error in other hyperfine coefficients, as they were calculated at the leading order from the well-known Breit-Pauli Hamiltonian~\cite{Bakalov06}, and higher-order corrections are of the order of 1~kHz for $E_1$~\cite{Korobov20} and smaller for other coefficients. In addition, experimental data on the rotational transition~\cite{Alighanbari20} has confirmed the theoretical values of these coefficients at the level of a few hundred Hz for the $(L=1,v=0)$ level. Concerning the theory of the spin-spin coefficients presented here, the main approximation, apart from neglecting the state dependence of higher-order QED corrections [Eq.~\ref{QED-ho}], consists in using the adiabatic approximation to calculate the $(Z\alpha)^2 E_F$ relativistic correction. The associated uncertainty can be estimated to be of relative order $(m/M)$, that is, smaller than 100~Hz. One additional feature of HD$^+$ (as compared to H$_2^+$) that is not taken into account in the adiabatic framework is the g/u symmetry breaking due to the mass asymmetry between proton and deuteron, which strongly affects rovibrational states close to the dissociation limit (see e.g.~\cite{Carrington89}). However, even the $(v=9,L=3)$ level is quite far from the dissociation limit, and the asymmetry of the wavefunction is still small (in the $10^{-3}$ range). In any case, a recalculation of the Breit correction in a full three-body approach would be highly desirable to test the accuracy of our results. Consideration of the state-dependent recoil correction of order~$(Z\alpha)^2(m/M)E_F$ might also be of interest.

In conclusion, we have presented a theory of higher-order corrections to the spin-spin scalar interaction in hydrogen molecular ions, and applied it to obtain improved values of the corresponding hyperfine coefficients for a range of rovibrational states in H$_2^+$ and HD$^+$. While the agreement with experimental data is excellent in H$_2^+$, a substantial discrepancy is observed in HD$^+$. It is currently unexplained, and will be the object of further investigations.

\begin{table} [t]
\small
\begin{tabular}{@{\hspace{2mm}}c@{\hspace{2mm}}c@{\hspace{4mm}}c@{\hspace{4mm}}c@{\hspace{10mm}}c
                @{\hspace{2mm}}c@{\hspace{4mm}}c@{\hspace{4mm}}c@{\hspace{2mm}}}
\hline \hline
  $L$  &  $v$  &  $E_4$  &  $E_5$  &  $L$  &  $v$  &  $E_4$  &  $E_5$ \\
\hline
  0  &  0  & 925.3942(9)  & 142.28756(8)  &  2  &  6  & 815.5646(8)  & 125.47914(7)  \\
  0  &  1  & 904.1471(8)  & 139.03010(8)  &  2  &  7  & 801.7350(7)  & 123.37411(7)  \\
  0  &  2  & 884.2889(8)  & 135.98714(8)  &  2  &  8  & 788.9278(7)  & 121.43086(7)  \\
  0  &  3  & 865.7411(8)  & 133.14692(8)  &  2  &  9  & 777.1025(7)  & 119.64475(7)  \\
  0  &  4  & 848.4337(8)  & 130.49899(8)  &  2  & 10  & 766.2212(7)  & 118.01236(7)  \\
  0  &  5  & 832.3039(8)  & 128.03411(8)  &  3  &  0  & 920.4800(9)  & 141.53307(8)  \\
  0  &  6  & 817.2953(8)  & 125.74423(7)  &  3  &  1  & 899.5058(8)  & 138.31764(8)  \\
  0  &  7  & 803.3575(7)  & 123.62237(7)  &  3  &  2  & 879.9078(8)  & 135.31479(8)  \\
  0  &  8  & 790.4451(7)  & 121.66266(7)  &  3  &  3  & 861.6089(8)  & 132.51296(8)  \\
  0  &  9  & 778.5169(7)  & 119.86036(7)  &  3  &  4  & 844.5404(8)  & 129.90192(8)  \\
  0  & 10  & 767.5349(7)  & 118.21194(7)  &  3  &  5  & 828.6406(8)  & 127.47263(8)  \\
  1  &  0  & 924.5677(9)  & 142.16067(8)  &  3  &  6  & 813.8543(8)  & 125.21721(7)  \\
  1  &  1  & 903.3665(8)  & 138.91027(8)  &  3  &  7  & 800.1320(7)  & 123.12888(7)  \\
  1  &  2  & 883.5519(8)  & 135.87404(8)  &  3  &  8  & 787.4294(7)  & 121.20197(7)  \\
  1  &  3  & 865.0459(8)  & 133.04026(8)  &  3  &  9  & 775.7061(7)  & 119.43193(7)  \\
  1  &  4  & 847.7786(8)  & 130.39852(8)  &  3  & 10  & 764.9249(7)  & 117.81546(7)  \\
  1  &  5  & 831.6874(8)  & 127.93962(8)  &  4  &  0  & 917.2621(9)  & 141.03904(8)  \\
  1  &  6  & 816.7161(8)  & 125.65551(7)  &  4  &  1  & 896.4674(8)  & 137.85123(8)  \\
  1  &  7  & 802.8145(7)  & 123.53928(7)  &  4  &  2  & 877.0404(8)  & 134.87475(8)  \\
  1  &  8  & 789.9372(7)  & 121.58507(7)  &  4  &  3  & 858.9052(8)  & 132.09817(8)  \\
  1  &  9  & 778.0434(7)  & 119.78818(7)  &  4  &  4  & 841.9938(8)  & 129.51139(8)  \\
  1  & 10  & 767.0951(7)  & 118.14511(7)  &  4  &  5  & 826.2452(8)  & 127.10551(8)  \\
  2  &  0  & 922.9238(9)  & 141.90827(8)  &  4  &  6  & 811.6051(8)  & 124.87277(7)  \\
  2  &  1  & 901.8138(8)  & 138.67192(8)  &  4  &  7  & 798.0247(7)  & 122.80652(7)  \\
  2  &  2  & 882.0862(8)  & 135.64910(8)  &  4  &  8  & 785.4602(7)  & 120.90121(7)  \\
  2  &  3  & 863.6634(8)  & 132.82815(8)  &  4  &  9  & 773.8718(7)  & 119.15243(7)  \\
  2  &  4  & 846.4759(8)  & 130.19874(8)  &  4  & 10  & 763.2228(7)  & 117.55703(7)  \\
  2  &  5  & 830.4616(8)  & 127.75173(8)  &     &     &              &               \\
\hline \hline
\end{tabular}
\caption{Values (in MHz) of the spin-spin scalar interaction coefficients $E_4$ and $E_5$ for rovibrational states $(L,v)$ of HD$^+$. \label{table-E4-E5}}
\end{table}

\appendix

\section{Relations between divergent matrix elements} \label{div-op}

In this Appendix, we will assume that the Coulomb potential $V$ is regularized in some way, and that the charge distribution $\rho$ ($4\pi\rho = \Delta V$) is a smooth function of space variables. We recall that the brackets denote an expectation value over the nonrelativistic wave function $\Psi_0$; the nonrelativistic energy is denoted by $E_0$. Other relevant definitions are given right after Eq.~(\ref{Heff6}).

The divergent terms that we want to transform are $\langle V_1 p^4 \rangle$ and $\langle V_1 p^2 V \rangle$, which appear in the second-order term, Eq.~(\ref{2nd-order-div}), and $\langle \rho_1 p^2 \rangle$, appearing in the first-order term, Eq.~(\ref{Heff6}). Using the relationship $p^2\Psi_0\!=\!2m(E_0\!-\!V)\Psi_0$, one obtains
\begin{subequations}
\begin{equation}
\langle V_1 p^4 \rangle = -2 m \langle V_1 p^2 V \rangle - 4 m^2 E_0 \langle V_1 V \rangle + 4m^2 E_0^2 \langle V_1 \rangle \label{V1p4}
\end{equation}
\begin{equation}\label{rho1p2}
\begin{array}{@{}l}\displaystyle
\langle \rho_1 p^2 \rangle = 2mE_0 \langle \rho_1 \rangle - 2m \langle V \rho_1 \rangle
\\[3mm]\hspace{10mm}\displaystyle
  = 2mE_0 \langle \rho_1 \rangle
  - m \left( \langle V_1 \rho + V \rho_1 \rangle + \langle V_2 \rho_1 - V_1 \rho_2 \rangle \right)
\end{array}
\end{equation}
\end{subequations}
Using commutation relations and integration by parts one can obtain the following relationships:
\begin{subequations}
\begin{equation}\label{B1}
\left\langle V_1p^2V \right\rangle =
   \left\langle V_1Vp^2 \right\rangle
   -2\pi\left\langle V_1\rho\!+\!V\rho_1 \right\rangle
   +\left\langle
      (V_1\boldsymbol{\mathcal{E}}\!+\!V\boldsymbol{\mathcal{E}}_1)
                                               \!\cdot\! \boldsymbol{\nabla}
   \right\rangle,
\end{equation}
\begin{equation}\label{B2}
\left\langle V_1p^2V \right\rangle =
   \left\langle
      \boldsymbol{\mathcal{E}}_1 \!\cdot\! \boldsymbol{\mathcal{E}}
   \right\rangle
   -\left\langle
      (V_1\boldsymbol{\mathcal{E}}\!+\!V\boldsymbol{\mathcal{E}}_1)
                                               \!\cdot\! \boldsymbol{\nabla}
   \right\rangle
   +\left\langle \mathbf{p}V_1V\mathbf{p} \right\rangle,
\end{equation}
\begin{equation}\label{B3}
\vrule width 0pt height 12pt
2\pi\bigl\langle V_1\rho\!+\!V\rho_1 \bigr\rangle =
   -\left\langle
      \boldsymbol{\mathcal{E}}_1 \!\cdot\! \boldsymbol{\mathcal{E}}
   \right\rangle
   +\left\langle
      (V_1\boldsymbol{\mathcal{E}}\!+\!V\boldsymbol{\mathcal{E}}_1)
                                               \!\cdot\! \boldsymbol{\nabla}
   \right\rangle.
\end{equation}
\end{subequations}
Subtracting Eq.~(\ref{B3}) from Eq.~(\ref{B1}), and using again $p^2 \psi_0 = 2m(E_0 - V) \psi_0$ in the second line, we obtain a suitable expression for the first required expectation value:
\begin{subequations}
\begin{eqnarray}\label{B4}
\left\langle V_1p^2V \right\rangle &=&
   \left\langle
      \boldsymbol{\mathcal{E}}_1 \!\cdot\! \boldsymbol{\mathcal{E}}
   \right\rangle
   +\left\langle V_1Vp^2 \right\rangle \nonumber \\
   &=&
   \left\langle
      \boldsymbol{\mathcal{E}}_1 \!\cdot\! \boldsymbol{\mathcal{E}}
   \right\rangle
   -2m\left\langle V_1V^2 \right\rangle
   +2mE_0\left\langle V_1V \right\rangle.
\end{eqnarray}
Adding up (\ref{B2}) and (\ref{B3}) and taking into account (\ref{B4}), we arrive at:
\begin{equation}\label{B5}
2\pi\bigl\langle V_1\rho\!+\!V\rho_1 \bigr\rangle =
   -\left\langle
      \boldsymbol{\mathcal{E}}_1\boldsymbol{\mathcal{E}}
   \right\rangle
   +2m\left\langle V_1V^2 \right\rangle
   -2mE_0\left\langle V_1V \right\rangle
   +\left\langle \mathbf{p}V_1V\mathbf{p} \right\rangle.
\end{equation}
\end{subequations}
Finally, using~(\ref{V1p4}) and ~(\ref{rho1p2}) we find the following expressions for the other two expectation values:
\begin{subequations}
\begin{equation}\label{c3}
\left\langle V_1p^4 \right\rangle =
   -2m\left\langle
      \boldsymbol{\mathcal{E}}_1\boldsymbol{\mathcal{E}}
   \right\rangle
   +4m^2\left\langle V_1V^2 \right\rangle
   -8m^2E_0\left\langle V_1V \right\rangle
   +4m^2E_0^2\left\langle V_1 \right\rangle,
\end{equation}
\vspace*{-4mm}
\begin{equation}\label{c4}
\begin{array}{@{}l}
4\pi\left\langle \rho_1p^2 \right\rangle =
   2m\left\langle
      \boldsymbol{\mathcal{E}}_1\boldsymbol{\mathcal{E}}
   \right\rangle
   -4m^2\left\langle V_1V^2 \right\rangle
   +4m^2E_0\left\langle V_1V \right\rangle
   -2m\left\langle \mathbf{p}V_1V\mathbf{p} \right\rangle
\\[2mm]\hspace{40mm}
   -4\pi m\left\langle V_2\rho_1\!-\!V_1\rho_2 \right\rangle
   +8\pi mE_0\left\langle \rho_1 \right\rangle.
\end{array}
\end{equation}
\end{subequations}


\begin{thebibliography}{99}
\bibitem{Wing76} W.H. Wing, G.A. Ruff, W.E. Lamb, Jr., and J.J. Spezeski,
   Observation of the Infrared Spectrum of the Hydrogen Molecular Ion HD$^+$,
   Phys. Rev. Lett.~\textbf{36}, 1488 (1976).

\bibitem{Alighanbari20} S. Alighanbari, G.S. Giri, F.L. Constantin, V.I. Korobov, and S. Schiller,
   Precise test of quantum electrodynamics and determination of fundamental constants with HD$^+$ ions,
   Nature~\textbf{581}, 152 (2020).

\bibitem{Patra20} Sayan Patra, M. Germann, J.-Ph. Karr, M. Haidar, L. Hilico, V.I. Korobov, F.M.J. Cozijn,
   K.S.E. Eikema, W. Ubachs, and J.C.J. Koelemeij,
   Proton-electron mass ratio from laser spectroscopy of HD$^+$ at the part-per-trillion level,
   Science~\textbf{369}, 1238 (2020).

\bibitem{Korobov17} V.I. Korobov, L. Hilico, and J.-Ph. Karr,
   Fundamental Transitions and Ionization Energies of the Hydrogen Molecular Ions with Few ppt Uncertainty,
   Phys. Rev. Lett.~\textbf{118}, 233001 (2017).

\bibitem{Bakalov06} D. Bakalov, V.I. Korobov, and S. Schiller,
   High-Precision Calculation of the Hyperfine Structure of the HD$^+$ Ion,
   Phys. Rev. Lett.~\textbf{97}, 243001 (2006).

\bibitem{Korobov09} V.I. Korobov, L. Hilico, and J.-Ph. Karr,
   Relativistic corrections of $m\alpha^6(m/M)$ order to the hyperfine structure of the H$_2^+$ molecular ion,
   Phys. Rev. A~\textbf{79}, 012501 (2009).

\bibitem{Korobov16} V.I. Korobov, J.C.J. Koelemeij, L. Hilico, and J.-Ph. Karr,
   Theoretical Hyperfine Structure of the Molecular Hydrogen Ion at the 1 ppm Level,
   Phys. Rev. Lett.~\textbf{116}, 053003 (2016).

\bibitem{Alighanbari18} S. Alighanbari, M.G. Hansen, V.I. Korobov, and S. Schiller,
   Rotational spectroscopy of cold and trapped molecular ions in the Lamb-Dicke regime,
   Nature Phys.~\textbf{14}, 555 (2018).

\bibitem{Korobov20} V.I. Korobov, J.-Ph. Karr, M. Haidar, and Z.-X. Zhong,
   Hyperfine structure in the H$_2^+$ and HD$^+$ molecular ions at order $m\alpha^6$,
   Phys. Rev. A~\textbf{102}, 022804 (2020).

\bibitem{Bethe77} H.A. Bethe and E.E. Salpeter,
   \textit{Quantum mechanics of one- and two-electron atoms}, Plenum Publishing Co., New York, 1977.

\bibitem{Sapirstein90} J.R. Sapirstein and D.R. Yennie, in: T. Kinoshita (Ed.),
   Q\textit{uantum Electrodynamics}, World Scientific, Singapore, 1990.

\bibitem{Kinoshita96} T. Kinoshita and M. Nio,
   Radiative corrections to the muonium hyperfine structure: The $\alpha^2(Z\alpha)$ correction,
   Phys. Rev. D~\textbf{53}, 4909 (1996).

\bibitem{Eides07} M.I. Eides, H. Grotch, and V.A. Shelyuto,
   \textit{Theory of Light Hydrogenic Bound States},
   Springer Tracts in Modern Physics~\textbf{222} (Springer, Berlin, 2007).

\bibitem{Mondejar10} J. Mond\'ejar, J. H. Piclum, and A. Czarnecki,
   Radiative-nonrecoil corrections of order $\alpha^2(Z\alpha)E_F$ to the hyperfine splitting of muonium,
   Phys. Rev. A~\textbf{81}, 062511 (2010).

\bibitem{Mohr16} P.J. Mohr, D.B. Newell, and B.N. Taylor,
   CODATA recommended values of the fundamental physical constants: 2014,
   Rev. Mod. Phys.~\textbf{88}, 035009 (2016).

\bibitem{Yerokhin08} V.A. Yerokhin and U.D. Jentschura,
   Electron Self-Energy in the Presence of a Magnetic Field: Hyperfine Splitting and $g$ Factor,
   Phys. Rev. Lett.~\textbf{100}, 163001 (2008).

\bibitem{Yerokhin10} V.A. Yerokhin and U.D. Jentschura,
   Self-energy correction to the hyperfine splitting and the electron $g$ factor in hydrogenlike ions,
   Phys. Rev. A.~\textbf{81}, 012502 (2010).

\bibitem{Sunnergren98} P. Sunnergren, H. Persson, S. Salomonson, S.M. Schneider, I. Lindgren, and G. Soff,
   Radiative corrections to the hyperfine-structure splitting of hydrogenlike systems,
   Phys. Rev. A~\textbf{58}, 1055 (1998).

\bibitem{Karshenboim00} S.G. Karshenboim, V.G. Ivanov, and V.M. Shabaev,
   Vacuum polarization in a hydrogen-like relativistic atom: hyperfine structure,
   Zh. Eksp. Teor. Fiz.~\textbf{117}, 67 [JETP~\textbf{90}, 59] (2000).

\bibitem{Volotka05} A.V. Volotka, V.M. Shabaev, G. Plunien, and G. Soff,
   Zemach and magnetic radius of the proton from the hyperfine splitting in hydrogen,
   Eur. Phys. J. D~\textbf{33}, 23 (2005).

\bibitem{Carlson08} C.E. Carlson, \textit{Proton Structure Corrections to Hydrogen Hyperfine Splitting},
   Lect. Notes in Phys.~\textbf{745}, 93 (Springer, 2008);
   C.E. Carlson, V. Nazaryan, and K. Griffioen,
   Proton structure corrections to electronic and muonic hydrogen hyperfine splitting,
   Phys. Rev. A~\textbf{78}, 022517 (2008).

\bibitem{Zemach56} A.C. Zemach,
   Proton structure and the hyperfine shift in hydrogen,
   Phys. Rev.~\textbf{104}, 1771 (1956).

\bibitem{Karshenboim97} S.G. Karshenboim,
   Nuclear structure-dependent radiative corrections to the hydrogen hyperfine splitting,
   Phys. Lett. A~\textbf{225}, 97 (1997).

\bibitem{Arnowitt53} R. Arnowitt,
   The Hyperfine Structure of Hydrogen,
   Phys. Rev.~\textbf{92}, 1002 (1953).

\bibitem{Newcomb55} W.A. Newcomb and E.E. Salpeter,
   Mass Corrections to the Hyperfine Structure in Hydrogen,
   Phys. Rev.~\textbf{97}, 1146 (1955).

\bibitem{Bodwin88} G.T. Bodwin and D.R. Yennie,
   Some recoil corrections to the hydrogen hyperfine splitting,
   Phys. Rev. D~\textbf{37}, 498 (1988).

\bibitem{Faustov02} R.N. Faustov and A.P. Martynenko,
   Proton polarizability contribution to hydrogen hyperfine splitting,
   Eur. Phys. J. C~\textbf{24}, 281 (2002).

\bibitem{Friar05} J.L. Friar and G.L. Payne,
   Nuclear corrections to hyperfine structure in light hydrogenic atoms,
   Phys. Rev. C~\textbf{72}, 014002 (2005).

\bibitem{Friar04} J.L. Friar and I. Sick,
   Zemach moments for hydrogen and deuterium,
   Phys. Lett. B~\textbf{579}, 285 (2004).

\bibitem{Beg74} M.A. Beg and G. Feinberg,
   Exotic Interactions of Charged Leptons,
   Phys. Rev. Lett.~\textbf{33}, 606 (1974); \textbf{35}, 130(E) (1975).

\bibitem{Karshenboim05} S.G. Karshenboim,
   Precision physics of simple atoms: QED tests, nuclear structure and fundamental constants,
   Phys. Rep.~\textbf{422}, 1 (2005).

\bibitem{Wineland72} D.J. Wineland and N.F. Ramsey,
   Atomic Deuterium Maser,
   Phys. Rev.~\textbf{5}, 821 (1972).

\bibitem{Aznabayev19} D.T. Aznabayev, A.K. Bekbaev, and V.I. Korobov,
   Leading-order relativistic corrections to the rovibrational spectrum of H$_2^+$ and HD$^+$ molecular ions,
   Phys. Rev. A~\textbf{99}, 012501 (2019).

\bibitem{Wolniewicz80} L. Wolniewicz and J.D. Poll,
   The vibration-rotational energies of the hydrogen molecular ion HD$^+$,
   J. Chem. Phys.~\textbf{73}, 6225 (1980).

\bibitem{Korobov13} V.I. Korobov, L. Hilico, and J.-Ph. Karr,
   Calculation of the relativistic Bethe logarithm in the two-center problem,
   Phys. Rev. A~\textbf{87}, 062506 (2013).

\bibitem{Jefferts69} K.B. Jefferts,
   Hyperfine structure in the molecular ion H$_2^+$,
   Phys. Rev. Lett.~\textbf{23}, 1476 (1969).

\bibitem{Zhong15} Z.-X.~Zhong, X.~Tong, Z.-C.~Yan, and T.-Y.~Shi,
   High-precision spectroscopy of hydrogen molecular ions,
   Chin. Phys. B~\textbf{24}, 053102 (2015).

\bibitem{Carrington89} A. Carrington, I.R. McNab, and C.A. Montgomerie,
   Spectroscopy of the hydrogen molecular ion,
   J. Phys. B~\textbf{22}, 3551 (1989).

\end{thebibliography}
\end{document}